\documentclass[twocolumn]{aastex631}
\usepackage[mathlines]{lineno}
\usepackage{mathrsfs}
\usepackage{amsmath}

\newcommand{\gsim}{\raisebox{-0.13cm}{~\shortstack{$>$ \\[-0.07cm]
      $\sim$}}~} 
\shorttitle{Thermal and Abundance Characteristics in a Flare}
\shortauthors{Xie et al.}
\graphicspath{{./}{figures/}}
\begin{document}
\title{Disentangling Thermal and Abundance Characteristics in a Solar Flare Using SDO/AIA, Hinode/XRT, and MinXSS-1 Observations}

\author[0009-0007-0582-7807]{Xiaoyan Xie}\thanks{E-mail: xiaoyan.xie@cfa.harvard.edu}
\affiliation{Harvard-Smithsonian Center for Astrophysics, 60 Garden Street, Cambridge, MA 02138, USA}
\author{Anna Chen}
\affiliation{Bowdoin College, 255 Maine Street, Brunswick, ME 04011 }
\affiliation{Harvard-Smithsonian Center for Astrophysics, 60 Garden Street, Cambridge, MA 02138, USA}

\author[0000-0001-5243-7659]{Crisel Suarez}
\affiliation{Vanderbilt University, 2201 West End Ave., Nashville, TN 37235, USA}
\affiliation{Harvard-Smithsonian Center for Astrophysics, 60 Garden Street, Cambridge, MA 02138, USA}
\author{Katharine K. Reeves}
\affiliation{Harvard-Smithsonian Center for Astrophysics, 60 Garden Street, Cambridge, MA 02138, USA}

\author{Soumya Roy}
\affiliation{Inter-University Centre for Astronomy and Astrophysics, Ganeshkhind, Pune, 411007, Maharashtra, India}
\affiliation{Harvard-Smithsonian Center for Astrophysics, 60 Garden Street, Cambridge, MA 02138, USA}

\author{Christopher S. Moore}
\affiliation{Harvard-Smithsonian Center for Astrophysics, 60 Garden Street, Cambridge, MA 02138, USA}

\author{Nishu Karna}
\affiliation{Harvard-Smithsonian Center for Astrophysics, 60 Garden Street, Cambridge, MA 02138, USA}

\author{Joy Velasquez}
\affiliation{Harvard-Smithsonian Center for Astrophysics, 60 Garden Street, Cambridge, MA 02138, USA}



\begin{abstract}
We investigate the thermal properties of a solar flare by the observations of soft X-ray Telescope (XRT) on board Hinode and the Atmospheric Image Assembly on board the Solar Dynamics Observatory (SDO/AIA).  Our analysis reveals a tenuous but hot plasma cloud located above the flare loops. The flare loops and  plasma cloud exhibit independent temperature profiles as a function of time, implying different heating mechanisms are present in these structures. We interpret the plasma cloud as an eruptive filament directly heated in the corona via magnetic reconnection that occurs when a rising filament interacts with this eruptive filament. Previous disk-integrated solar soft X-ray spectral measurements of this flare using the Miniature X-ray Solar Spectrometer CubeSat-1 (MinXSS-1) inferred coronal abundances at the flare peak. However, MinXSS synthetic spectra from AIA observations show that the plasma cloud is too tenuous to be detected in MinXSS‑1 and thus the coronal abundances of low FIP elements must be in emission from the flare loops. 
Furthermore, we demonstrate the non-negligible differences in differential emission measures (DEMs) between coronal and photospheric abundance models. These results  highlight the importance of instruments capable of simultaneous spectral and spatial diagnostics during large-scale solar events. Such coordinated measurements would enable more accurate thermal and compositional analyses and deeper insight into the underlying physical mechanisms.
\end{abstract}


\section{Introduction} \label{sec:intro}
Solar flares are one of the most energetic activities in the solar system \citep{2000JGR...10523153F,2011SSRv..159...19F}. Magnetic reconnection, where the magnetic energy converts to kinetic energy and plasma heating with the breaking and rejoining of magnetic field lines, is the key driver for solar flares \citep{2002A&ARv..10..313P,2003NewAR..47...53L,2006SSRv..123..251F,2011LRSP....8....6S,2017A&A...604L...7M,2022MNRAS.509..406X}. It is now generally believed that during flares, when energy is deposited into the lower atmosphere from solar corona, dense photospheric/chromospheric  plasma is heated and evaporated filling the loops \citep{1985ApJ...289..425F,1987ApJ...317..502F,1991plsc.book.....B,2002ApJ...578..590R,2009ApJ...699..968M,2014ApJ...795...10L,2014LRSP...11....4R,2016ApJ...820...14Q}. This process is usually called chromospheric evaporation. 

Compared to photospheric abundances, the low first ionization potential (FIP $\lesssim$ 10 eV) elements in the corona are often enriched, while the high-FIP elements remains almost unchanged \citep{1992ApJS...81..387F,2014ApJ...786L...2W}, in the corona. The fractionation of low and high FIP elements are assumed to happen in the chromosphere \citep{2004ApJ...614.1063L,2015LRSP...12....2L}. The difference between the distributions of FIP elemental abundances in the corona and photosphere enables the diagnosis of elemental abundances from spectra to provide clues about the heating processes during solar eruptions. 

By comparing the relative intensities of Fe emission lines and continuum emission from the observations of EUV Variability Experiment (EVE)
on the Solar Dynamics Observatory (SDO), \citet{2014ApJ...786L...2W} calculated the FIP bias $f$ (the enhancement of the plasma relative to the composition of the photosphere) in 21 flares. The mean FIP bias $f$ = 1.17 $\pm$ 0.22, which is close to photospheric composition, indicating that plasma evaporation occurs deep in the chromosphere below the layer where elemental fraction happens during the flares. The decreasing trend of the abundance of the low-FIP elements during the flare onset phase is also shown in \citet{2021ApJ...920....4M,2023ApJ...957...14S,2024ApJ...966..198T}, highlighting the occurrence of chromospheric evaporation during flares. 

In the case of direct plasma heating in the corona, the elemental abundances may tell a different story. The thermal diagnostics of the plasma sheet associated with an eruptive X8.3 flare shows that the temperature of the plasma sheet is $\gsim$10~MK \citep{2018ApJ...866...64C,2018ApJ...854..122W,2024ApJ...972..164G,2024FrASS..1183746X}. However, the plasma sheet is also observed in 171 \AA\ and 211 \AA\ channels of Atmospheric Image Assembly on board the Solar Dynamics Observatory (SDO/AIA, \citealp{2012SoPh..275...17L}), lines that do not have a strong contribution from emission lines at the high temperatures of the plasma sheet \citep{2010A&A...521A..21O}, indicating that the emission of the plasma sheet observed in these two channels mainly comes from thermal bremsstrahlung continuum. \citet{2018ApJ...854..122W} compared emission measure in 171 \AA\ and 211 \AA\ channels with EIS wavelengths and AIA channels that are dominated by Fe line emission,  concluding that the plasma within the plasma sheet contains coronal composition, and the plasma sheet is formed by the direct heating of the plasma in the corona.  

The studies of abundances by \citet{2018ApJ...853..178D} and \citet{2024A&A...691A..95T} using the same event as \citet{2018ApJ...854..122W} complicate the standard picture that flare loops should have photospheric abundances.  These studies find that there is a high FIP bias in the  flare loop tops and a FIP bias of $\sim$1 around footpoints in this event.  \citet{2024A&A...691A..95T} explain the variation as due to the confinement of high FIP bias plasma downflows from the plasma sheet to the loop tops and chromospheric evaporation that fills the loop footpoints from below with low FIP bias plasma. And \citet{2018ApJ...853..178D} pointed out another possibility for the variation, which is that the plasma in the loop tops are a blend of loops or magnetic strands along the line of sight, and some strands have been in the corona long enough for the abundances to become coronal.

In 21 flares that were analyzed using disk-integrated solar spectrum of Miniature X-ray Solar Spectrometer CubeSat-1 (MinXSS-1) CubeSat Mission \citep{2016JSpRo..53..328M,2017ApJ...835..122W,2018SoPh..293...21M}, \citet{2023ApJ...957...14S} found that the abundance evolution in most cases is consistent with the chromospheric evaporation model where abundances of low-FIP elements, Fe, Ca, Si, S, and Mg are depleted from coronal toward photospheric values during the peak of the flares. In contrast to the general scenario, the elemental abundances of the flare 2016-07-21 M1.0 flare (Figure~10 in \citealp{2023ApJ...957...14S}) shows an ``anomalous" elemental abundance evolution. The abundances of low-FIP elements increase from the photospheric value in the impulsive phase to coronal values around the flare peak (this trend is remarkably shown in the evolution of Si abundance, see second row of Figure 10 in \citealp{2023ApJ...957...14S} and right panel Figure~\ref{goes}), indicating that the heating of coronal plasma is dominant around the flare peak. The temperature diagnostics of 2016-07-21 flare in  \citet{2023ApJ...957...14S} show a long period ($\sim$1 hour) of high temperatures ($\gsim$ 7~MK) throughout the impulsive, peak, and decay phases of the flare.

In this analysis, we investigate the evolution of the thermal characteristics of the flare using observations from soft X-ray Telescope (XRT, \citealp{2007SoPh..243...63G}) on board Hinode and SDO/AIA to perform thermal diagnostics for the 2016-07-21 flare. With images of Hinode/XRT and SDO/AIA, we examine the plasma heating of the flare and the processes that could be responsible for the ``anomalous" evolution of the low-FIP elemental abundances observed in \citet{2023ApJ...957...14S}. We also compare the differential emission measures (DEMs) between implementing different abundance models. DEMs with additional XRT filters are obtained and compared with the AIA-only DEMs. In the next section, we will present the observations of the flare from Hinode/XRT, SDO/AIA, and MinXSS-1. In Section~\ref{data-method}, we will introduce data process and the methods of generating the temperature response and calculating the DEMs in this work. We will present the results and discuss our work in Sections~\ref{sec:results} and~\ref{sec:discussion}, respectively. 

\section{Observations}
\label{Observations}
\begin{figure}
\centering
\includegraphics[scale=0.33]{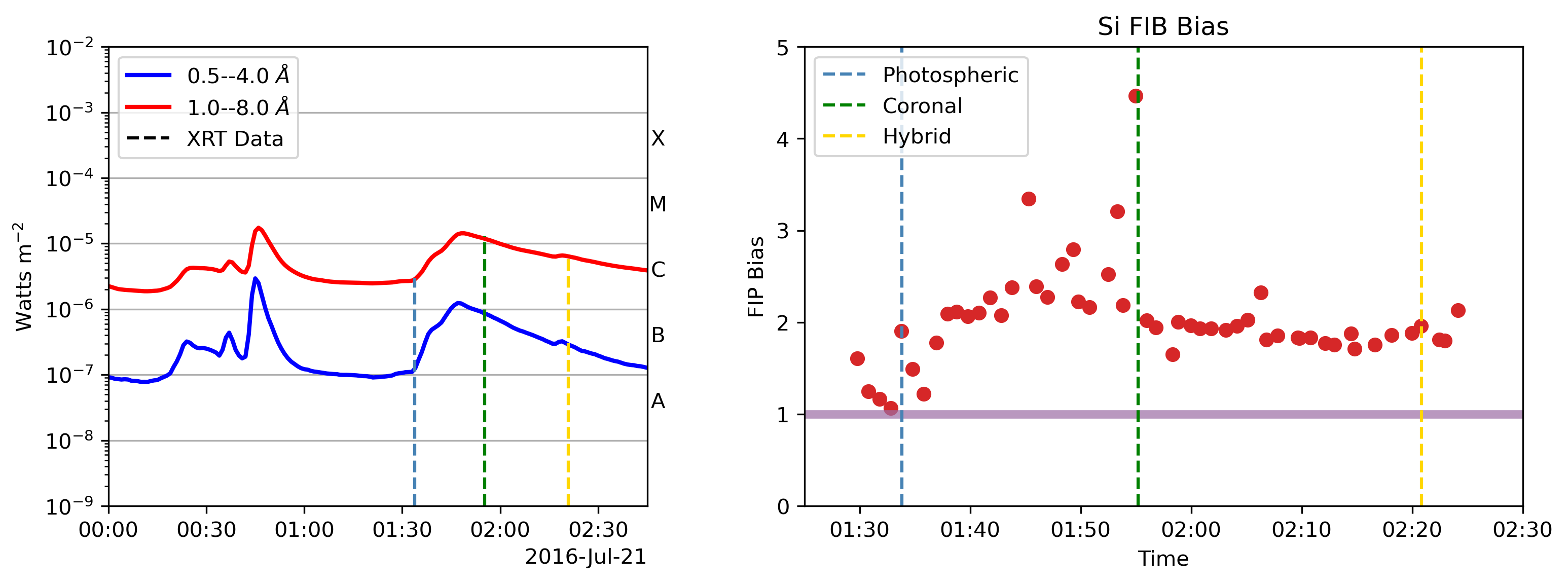}
\caption{Left: GOES flux around the flare; right: the evolution of Si during the flare (adapted from \citet{2023ApJ...957...14S}) where the purple bar indicates the photopheric value with FIP bias as 1. The dashed lines indicate the times with XRT observations, and the corresponding abundance models are labeled by different colors.}
\label{goes}
\end{figure}
\begin{figure}
\centering
\includegraphics[scale=0.45]{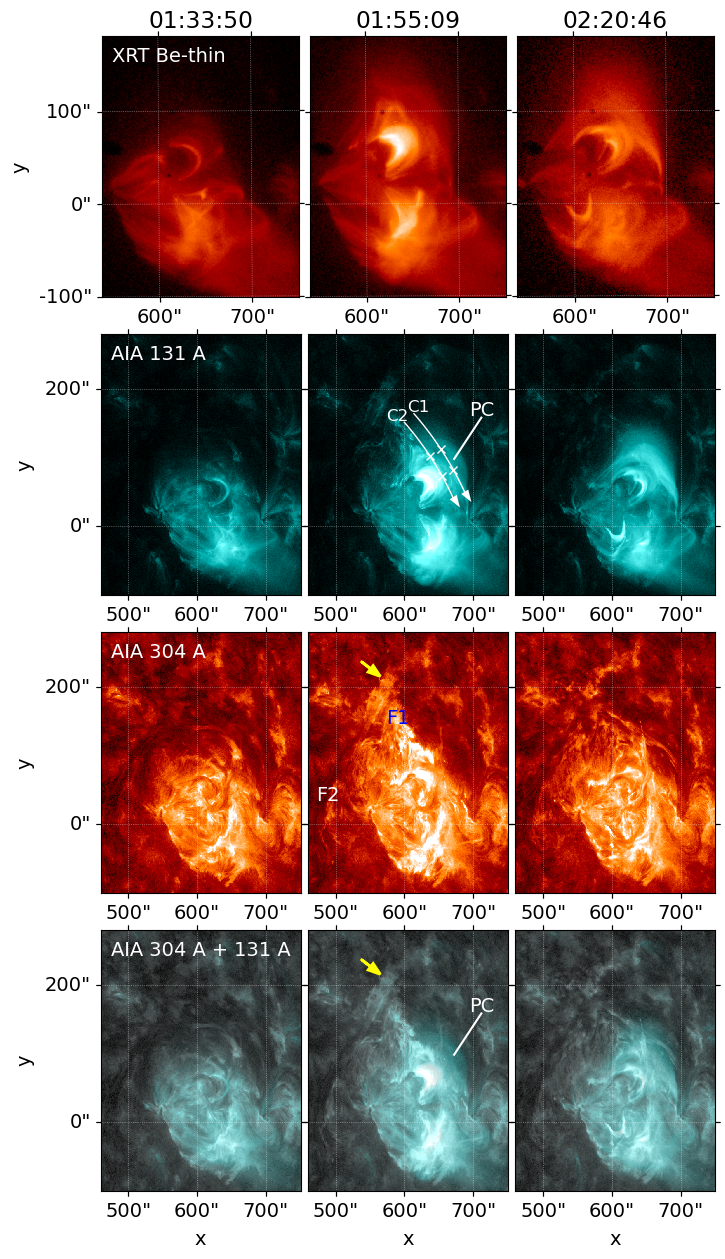}
\caption{Evolution of the flare and filament in XRT Be-thin (top row), AIA 131~\AA\ (second row), AIA 304~\AA\ (third row), and 304~\AA\ overlaid with 131~\AA\ (bottom row). Eruptive filament, pre-existing filament that crosses [470$^"$, 0] to [650$^{"}$, 250$^{"}$], and plasma cloud are indicated by F1, F2, and PC, respectively. The yellow arrows point out filament brightening where the eruptive filament F1 approaches the pre-existing filament F2. Curves C1, C2 indicate two paths for the position-vs.-time stackplots in Figures~\ref{aiastack} and~\ref{stackplot} where white arrows on the curves correspond to distance increment direction. The crosses marked in the curves correspond to dotted lines in the stackplots in Figures~\ref{aiastack} and~\ref{stackplot}.  Note that the plotting ranges in XRT and AIA data are different. The corresponding animation from 01:00 UT to 02:27 UT is available online.
}
\label{filament}
\end{figure}

On 2016 July 21, a flare of magnitude M1.0 erupted on the northwest of the Sun. It started at 01:34 UT, and the X-ray intensity achieved its peak at 01:49 UT (left panel in Figure~\ref{goes}). Hinode/XRT captures the onset of the impulsive phase, peak, and decay phase of the flare, and the times with the observations of XRT Be-med and Be-thin filters are indicated by the dashed lines in Figure~\ref{goes}. The background 1-8 \AA\ X-ray flux before the impulsive phase is on the level comparable with a C2 flare, which may be the remains of a previous flare with an X-ray class of M2 that peaked at 00:45 UT. After examining the XRT and AIA data, we noted that the previous M2 flare happened in the same active region as our studied flare. We see some formed flare loops in the southern hemisphere that come from the previous M2 flare at the beginning of our studied flare, as shown in 01:33:50 UT in Figure~\ref{filament}. 

The right panel of Figure~\ref{goes} shows that the Si abundance increases from a photospheric value in the impulsive phase to a coronal value in the flare peak. The Si abundance decays from the coronal value to a value between the coronal and photospheric values during the decay phase.

Figure~\ref{filament} shows the evolution of flare from the observations of XRT Be-thin and AIA 131 and 304 \AA\ channels. The flare loops with two major orientations brighten up from the flare impulsive phase are observed in XRT Be-thin and AIA 131 \AA. Note that the flare loops and the surrounding plasma in the southern southern part of the active region, which formed from the previous M2 flare, exist at the beginning of our studied flare. The flare loops in the southern hemisphere brighten up again during our studied flare.

After the flare impulsive phase, the filament flux rope (indicated by F1 in Figure~\ref{filament}) rises up, and it is most clearly seen in 304 \AA, as shown in the middle row of Figure~\ref{filament} (and its corresponding animation). We see the formation of ``plasma cloud" (the tenuous plasma above the northern loops, which is most obviously shown in XRT Be-thin and AIA 131~\AA, indicated by ``PC" in Figure~\ref{filament}) during the same period when the flux rope rises. Comparison between AIA 131 \AA\ and XRT Be-thin shows that XRT Be-thin observes more extensive plasma above the flare loops while AIA 131 \AA\ observes finer structure in the partially erupted flux rope (also see the animation of Figure~\ref{filament}) around [580", 120"]. In the the bottom row of Figure~\ref{filament}, where we overlay the AIA 131 \AA\ on AIA 304 \AA, the eruptive filament brightening in 304 \AA\ does not co-spatially overlap with the plasma cloud observed in 131 \AA. The formation of the plasma cloud in 131 \AA\ could be the heated eruptive flux rope that is hotter than the low temperature channel of 304~\AA. The disappearance of the filament in the GONG \citep{Harveyetal96} $H\alpha$ observations after the flare further confirms that the filament erupts during the flare.

The intersection of the pre-existing filament (indicated by F2 in Figure~\ref{filament}) that crosses [470$^"$, 0] to [650$^{"}$, 250$^{"}$] and the eruptive filament F1 brightens (indicated by a yellow arrow at 01:54:07 UT in Figure~\ref{filament}) during $\sim$ 01:52 UT to 02:20 UT. 
We note that there is plasma ejected from this intersection, as shown in the animation of Figure~\ref{filament}. 
\begin{figure}
\centering
\includegraphics[scale=0.38]{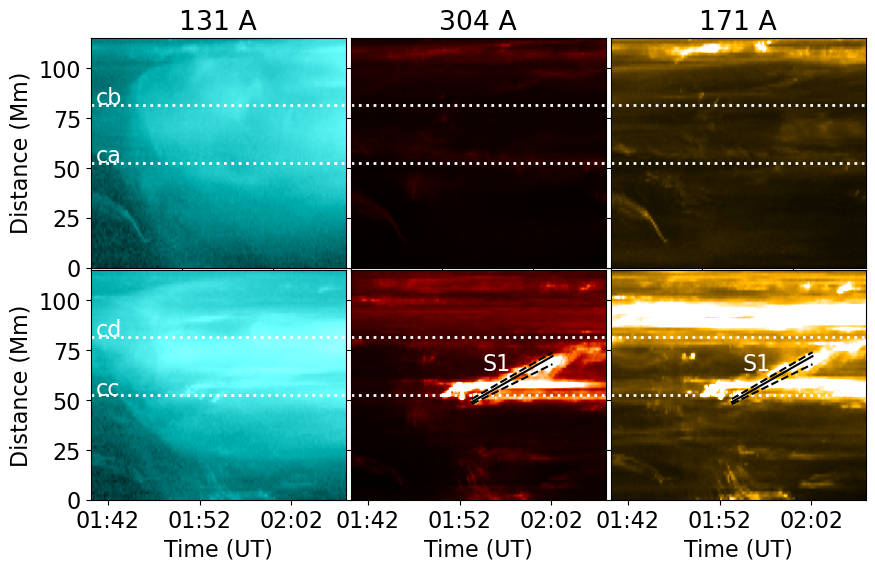}
\caption{The position-vs.-time stackplots of 131~\AA, 304~\AA, and 171~\AA\ distribution along the paths of C1 (top row) and C2 (bottom row) indicated in Figure~\ref{filament}. The distance increment corresponds to the arrow direction indicated on the paths C1 and C2. The dotted lines ca, cb, cc, and cd correspond to crosses indicated in Figure~\ref{filament}. S1 indicates the ejected outflows from brightening intersection of eruptive filament F1 and pre-existing filament F2. The fitted speed of the outflows, indicated by solid lines, is 44 km/s. The minimum and maximum fitted speeds, shown by dashed lines \citep{2015ApJ...807....7R}, are 38 km/s and 45 km/s, respectively.}
\label{aiastack}
\end{figure}

We perform position-vs.-time stackplots of 131~\AA, 304~\AA, and 171 \AA\ distributions along the paths of C1 and C2 that are indicated in Figure~\ref{filament}. These stackplots are shown in Figure~\ref{aiastack} with the top and bottom rows showing the distribution along C1 and C2, respectively. Path C1 mainly passes through the plasma cloud that is most obviously observed in 131~\AA, and path C2 mainly passes through part of the eruptive flux rope that is bright in 304 \AA. 

The formation of the plasma cloud, which can be seen in the stackplots of the high temperature response AIA 131 \AA\ channel, starts at $\sim$ 01:42 UT when the filament flux rope F1 starts rising up (see the animation of Figure~\ref{filament}). During 01:52 UT to 02:05 UT, C2 captures part of the ejected plasma from the brightening site of intersection of F1 and pre-existing filament F2, as indicated by slope S1 on the stackplots in 304 \AA\ and 171 \AA. The shift of brightening plasma in 171~\AA\ and 304~\AA\ stackplots has been reported in \citet{2014ApJ...788...85V} as direct evidence of magnetic reconnection and
resulting in in-situ heating. Compared to their analysis, we have not captured evident magnetic reconnection inflows either from the stackplots with the paths parallel or perpendicular to path C1. One reason of this could be the 3D geometry of our rising flux rope, i.e., the flux rope rises along the dimension perpendicular to the x-y plane (see Figure~\ref{filament}), which makes the inflow velocity harder to capture in the plane of the sky. Even so, the plasma drift indicated by S1 is likely the manifestation of magnetic reconnection outflows. More evidence of plausible magnetic reconnection happening around the brightening filament area and the role magnetic reconnection plays in the heating of the plasma cloud is further discussed in later sections detailing the thermal diagnostics on the flare. The velocity of the brightening plasma drift in \citet{2014ApJ...788...85V} is $\sim$ 90 km s$^{-1}$, and they conclude this high speed is due to the combination of successive reconnections and the shift of the reconnection region. In our case, the estimated plane-of-sky speed of the plasma that corresponds with plasma drift indicated by S1 is 44$\pm$6 km/s. Several factors could contribute to the speed difference between \citet{2014ApJ...788...85V} and our results. The reconnection in \citet{2014ApJ...788...85V} is between the flux rope and surrounding magnetic structures, a pattern similar to that in \citet{2022ApJ...933..148C} and \citet{2024ApJ...977L..26C} while the reconnection in our case is between two filaments. The speed of plasma derived from stackplots is susceptible to the chosen paths. The path we chose is empirical, and it can generally capture the trajectories of motions of most outflows. Eventually, we have more projection effects than in \citet{2014ApJ...788...85V}. Using the de-projection approach described in Appendix A of \citet{2025arXiv250909944X}, we estimate the speed of outflows after de-projection is 66$\pm$9 km/s.

\section{Data and Methods}
\label{data-method} 
\subsection{Data Process}
For AIA data, we promote them from level 1 to level 1.5 by functions `update\_pointing' and `register' in aiapy.calibrate subpackages of aiapy \citep{Barnes2020}. The promotion updates the pointing keywords, removes the roll angle of the satellite, rescales the image to a resolution of 0.6$^{"}$/pixel, and aligns the center of the image with the center of the Sun. We obtain the degradation by `aiapy.calibrate.degradation' function. AIA data are normalized by dividing by the exposure time and further corrected by dividing by the degradation.

We download the XRT level 1 data from https://xrt.cfa.harvard.edu/level1/, which has corrected the roll angle and degradation and normalized the exposure time, and is ready for the use of science analysis. Since XRT and AIA images have different pixel scales, for DEM calculations that combine XRT filters with AIA channels, we need to adjust pixel scale of AIA to be the same as XRT. To do so, we use `resample' function to make AIA images have same pixel numbers as XRT along each dimension for the same region on the solar coordinate. Eventually, we overlay AIA images on XRT to check whether the images are co-aligned. We adjust the transparency of the images by `alpha' keyword on image visualization in Python and change the layer orders to make sure our eyes can identify the relative location change of half a pixel between XRT and AIA images by blinking images.

\subsection{Temperature Response and Differential Emission Measurement (DEM)}
In order to better understand the heating processes in this flare, we employ DEMs to probe the temperature structure in the flare. The principle behind DEMs calculation is that the observed flux $F_i$ for each bandpass of SDO/AIA EUV channels \citep{2012SoPh..275...41B,2012ApJ...761...62C} and Hinode/XRT soft X-ray channels can be expressed as 
\begin{equation}
   F_i = \int R_i(T) \times {\rm DEM}(T) dT
   \label{flux}
\end{equation}
Where $R_i$ is the temperature response function for $i$ bandpass in AIA or XRT, and the DEM is typically given by \citep{2012SoPh..275...41B,2012A&A...539A.146H} 
\begin{equation}
     {\rm DEM} (T) = n^2(T, z)\frac{dz}{dT}, 
     \label{eqdem}  
    \end{equation}
 with units of cm$^{-5}$ K$^{-1}$. In Eq.~\ref{eqdem}, $n$ is the number density and $dz$ is the length element along the line of sight (LOS).  The $\rm {DEM}(T)$ enables us to calculate an DEM-weighted average temperature 
\begin{equation}
   \overline{T}=\frac{\int_{T1} ^ {T2} T \times {\rm DEM}(T)dT}{\rm EM}, 
\end{equation}  
where the emission measure (EM) is given by
\begin{equation}
    {\rm EM} = \int_{T1} ^ {T2} {\rm DEM}(T) dT=\int_{z1} ^ {z2} n^2(z) dz.
    \label{eqem}
\end{equation}
\begin{figure}
\centering
\includegraphics[scale=0.4]{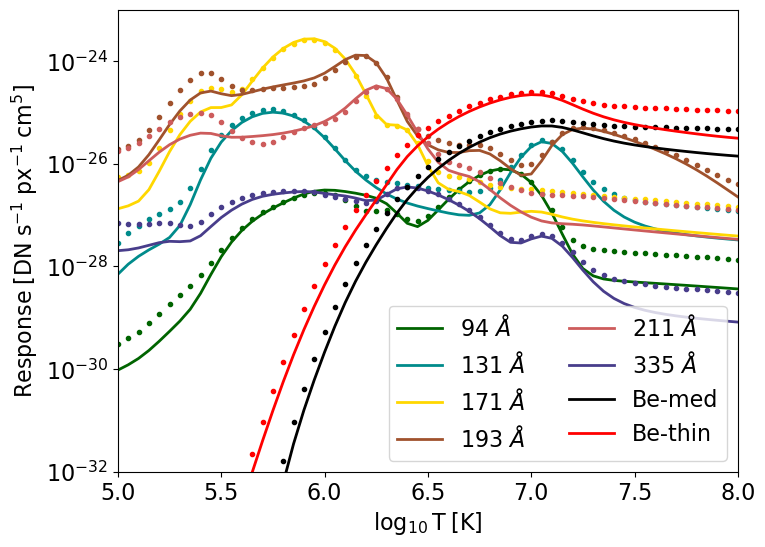}
\caption{SDO/AIA and Hinode/XRT response functions with coronal abundances (solid curves) and photospheric abundances (dotted lines, scaled by a factor of 3.98) for the date of 2016-07-21. Following the practice in \citet{2017ApJ...844..132W}, XRT response functions are further multiplied by a factor of 2.}
\label{tresponse}
\end{figure}
Equation~\ref{eqem} indicates that the distribution of EM can tell us the general distribution of the density of the structures along the LOS. We implement T1=$10^{5}$~K and T2=$10^{7.4}$~K in our calculation. The DEMs herein are recovered by the approach of regularized inversion developed by \citet{2012A&A...539A.146H} and \citet{2013A&A...553A..10H} (please refer to these two papers for further details).  We use six EUV channels (94 \AA, 131 \AA, 171 \AA, 193 \AA, 211 \AA, and 335 \AA) that are dominated by the emission from iron lines in SDO/AIA, and XRT Be-med and Be-thin broadband filters when available to derive the temperature. We use 40 temperature bins in the range of $10^{5}$ K to $10^{7.4}$ K with the interval $\Delta$log T evenly distributed. Figure~\ref{tresponse} shows the temperature response $R_i$ of AIA and XRT channels where solid curves indicate coronal abundances from \citet{1992ApJS...81..387F} and dotted curves indicate photospheric abundances from \citet{2009ARA&A..47..481A}, scaled by a factor of 3.98 (as the ratio of iron abundance from the corona to photosphere is 3.98. This follows the same practice as in Figure~A.2 of \citealp{2011A&A...535A..46D}). 

The temperature responses in Figure~\ref{tresponse} are generated by folding the effective areas with the isothermal spectra, as is the practice in \citet{2011A&A...535A..46D}. The isothermal spectra is obtained using the CHIANTI routine {\tt{isothermal.pro}} with the selection of abundance model and AIA effective areas obtained by {\tt{aia$\_$get$\_$response.pro}} (also see the sample code in the Appendix of \citealp{2011A&A...535A..46D}). The {\tt {isothermal.pro}} and {\tt{aia\_get\_response.pro} }are both available in the open SolarSoftware \citep[SSW;][]{1998Freeland} package. The XRT effective areas are obtained using the {\tt{EffectiveAreaFundamental.effective\_area()}} function in the {\tt{xrtpy.response}} module of the XRTpy package \citep{2024JOSS....9.6396V}. Since the DEM approach is to constrain the ill-posed inverse problem \citep{2012A&A...539A.146H,2013A&A...553A..10H} with the under-determined system of M $>$ N (M is the temperature bin number and N is the number of channels), including additional two XRT filters to six AIA channels can better constrain the DEM results \citep{2014ApJ...786...95H}. 

There are prior studies suggesting that the XRT temperature response functions are lower than the ones expected from DEM calculations using the data from other instruments \citep{2011ApJ...728...30T,2015ApJ...807..143C,2018ApJ...856L..17S}. It has been found that when combining XRT filters with AIA channels (and $NuSTAR$ spectra) to calculate DEM, multiplying XRT response function by a factor of 2 or 3 makes the predicted XRT intensities from AIA DEMs \citep{2015ApJ...806..232S} or $NuSTAR$ spectral fits \citep{2017ApJ...844..132W} closer to observed XRT intensities. After tests (see Figure~\ref{xrtfactors} and associated discussion in the Appendix), we found that, in our case, multiplying XRT response function by factor 2 makes the predicted XRT intensities from the AIA DEMs closer to XRT intensities from observation, and the results have a better agreement as indicated by smaller $\chi^2$ (also see \citealp{2017ApJ...844..132W}) from DEM results. We therefore multiply XRT response function by factor 2 for the case of combining XRT with AIA filters for calculating DEMs. 

\section{Results} 
\label{sec:results}
In this section, we present the thermal diagnostics of the flare. We first investigate plasma heating process during the flare by calculating DEMs using AIA data. Then we compare the DEMs between implementing different abundance models. Eventually, DEMs with additional XRT filters are obtained and compared with the AIA-only DEMs.

\subsection{Plasma heating during the flare} 
\begin{figure*}
\centering
\includegraphics[scale=0.7]{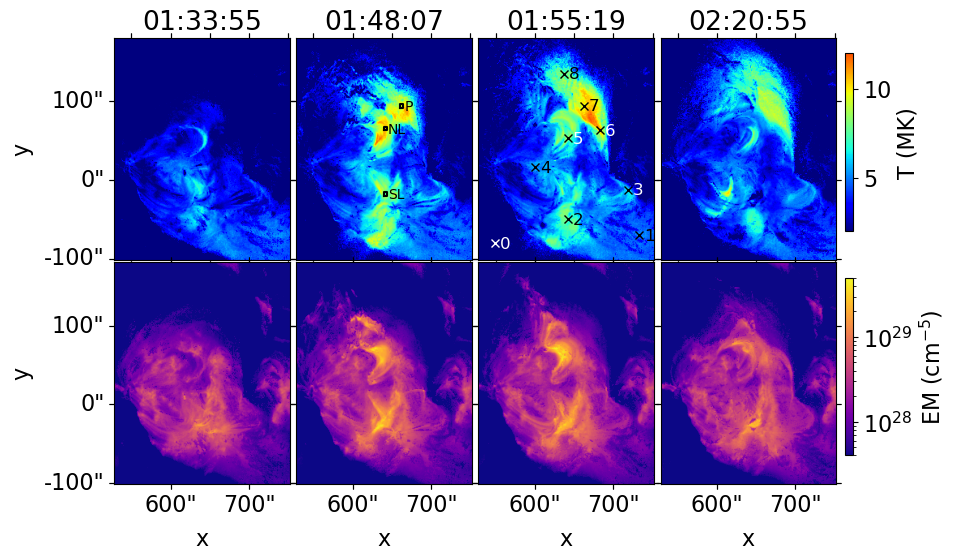}
\caption{Evolution of the temperature (top row) and EM (bottom row) of the flare from AIA observations. Boxes at 01:48:07 UT marks the locations of plasma cloud (P), northern loops (NL), and southern loops (SL) for plotting temperature as a function of time in Figure~\ref{tcurve}. Crosses mark the locations for checking DEM-temperature distributions at 01:55:19 in Figure.~\ref{peak_diffabun_curve}.}
\label{T_aia_evolution}
\end{figure*}

Figure~\ref{T_aia_evolution} shows the evolution of the temperature (top row) and EM (bottom row) of the flare from AIA observations. From 01:33:55 UT to 01:48:07 UT, the flare undergoes the impulsive phase to the flare peak, and \citet{2023ApJ...957...14S} shows a temperature increase during this time period for this event. Temperature and EM diagnostics in Figure~\ref{T_aia_evolution} show that both northern and southern flare loops with dense plasma accumulate and heat up rapidly, consistent with chromospheric evaporation where lower atmospheric plasma is heated and evaporated to fill the flare loops. 
We also note that a hot ``plasma cloud" forms and heats up quickly mostly above the northern flare loops and partially above the southern flare loops (around [640$^"$, -80$^"$]) during the same time period. After referring to bottom panels of Figure~\ref{filament} and its animation, we found that the hot plasma cloud is geometrically located above the bright eruptive filament F1 observed in 304 \AA\ and could be part of the eruptive flux rope that is heated to temperatures that are above the temperature response range of the low temperature channel of 304 \AA\ but within the temperature response range of high temperature channel of 131 \AA. The filament rises up and brightens up quickly in 304 \AA\ during the flare impulsive phase to the peak. The heated plasma cloud is more tenuous than the flare loops, as shown in EM distributions at the bottom of Figure~\ref{T_aia_evolution}. 
\begin{figure}
\centering
\includegraphics[scale=0.45]{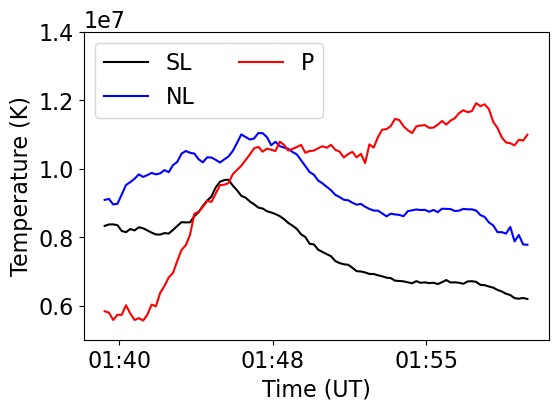}
\caption{Temperature as a function of time in the regions indicated by boxes of plasma cloud (P), northern loops (NL), and southern loops (SL) in Figure~\ref{T_aia_evolution}.}
\label{tcurve}
\end{figure}

From 01:48:07 to 01:55:19 UT, the flare undergoes the decay phase, and both northern and southern loops cool down. During the same time period, the plasma cloud above the northern flare loops continues to heat up and the heated plasma cloud becomes more spatially extensive, implying that there are different heating and cooling phases occurring in the flare loops and the plasma cloud. The heating of the plasma cloud is independent from that of the flare loops and the southern above-the-loop-top regions. This effect is more apparently shown in Figure~\ref{tcurve}, which plots the temperature as a function of time in several regions. In this plot, the northern and southern loops show heating/cooling simultaneously, but the hot plasma cloud above the northern flare loops displays independent heating/cooling behavior compared to the loops. At 02:20:55 UT, $\sim$ 47 minutes after the onset of the flare, both northern and southern flare loops cool down to the temperature close to the one before the onset of the flare. The plasma cloud cools down compared to $\sim$25 minutes ago but still maintains a relatively high temperature ($\sim$10 MK).  

\begin{figure}
\centering
\includegraphics[scale=0.45]{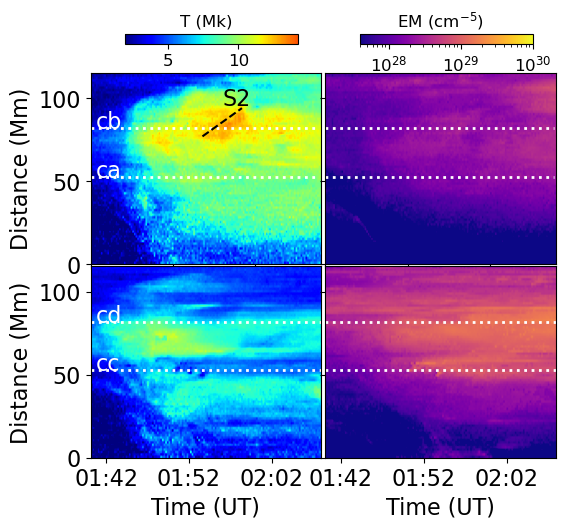}
\caption{The position-vs.-time stackplots of temperature and EM distributions along the paths of C1 (top row) and C2 (bottom row) indicated in Figure~\ref{filament}. The distance increment corresponds to the arrow direction indicated on the paths C1 and C2. The dotted lines ca, cb, cc, and cd,  correspond to crosses indicated in Figure~\ref{filament}. S2 indicates high temperature drift in the plasma cloud.}
\label{stackplot}
\end{figure}

Figures~\ref{T_aia_evolution} and~\ref{tcurve} show that the plasma cloud has the highest temperature of the flare and displays pronounced heating around the flare peak at 01:55:19 UT. However, Figures~\ref{T_aia_evolution} indicates that the EM of plasma cloud is at least one order of magnitude lower than that of the loops, which makes the detection of plasma cloud in disk-integrated spectra of MinXSS-1 questionable. We calculate the MinXSS-1 synthetic spectra from AIA observations (see Figure~\ref{syn} in the Appendix). Figure~\ref{syn} shows that the synthetic spectrum from the plasma cloud is about one order of magnitude lower than that of the loops, indicating that the plasma cloud is too tenuous to be detected in MinXSS‑1. This result suggests that the plasma cloud is not the source of the coronal abundances in the MinXSS observations.  Rather, the coronal abundances of the low FIP elements could come from flare loops \citep{2024A&A...691A..95T}. The existence of coronal abundances in flare loops is consistent with our DEM $\chi^2$ distributions in Figure~\ref{chisq}. It shows that loops with coronal abundances have generally lower $\chi^2$ than that with photospheric abundances. Another possibility is that the temperature sensitivity is different between MinXSS-1 and AIA so that the synthetic spectra can not fully represent the MinXSS-1 observations. Either cause is out the scope of current work, and it requires more future studies to investigate the mechanisms responsible for coronal abundances detected in the flare peak.

We perform position-vs.-time stackplots of temperature, and EM distributions along the paths of C1 and C2  indicated in Figure~\ref{filament}. The stackplots of 131 \AA, 304 \AA, and 171 \AA\ along paths C1 and C2 are shown in Figure~\ref{aiastack}. The plasma density along C2 is generally higher than along C1, as we see the EM in the bottom of Figure~\ref{stackplot} are higher than in the top. 

The plasma along C2 first shows high temperature when the filament flux rope F1 starts rising up at $\sim$ 01:42 UT (see the animation of Figure~\ref{filament}). The plasma cloud exhibits higher temperatures a little later (see temperature stackplots for C1 in top left of Figure~\ref{stackplot}), starting at  $\sim$ 01:44 UT, and the stackplot shows that there is prominent heating that becomes the dominant high temperature area after $\sim$ 01:50 UT. The formation of the high temperature plasma cloud can be recognized in the stackplots of the high temperature response AIA 131 \AA\ channel and can not be recognized in low-temperature channel of 304 \AA\ and 171 \AA~(Figure~\ref{aiastack}). Since the plasma heating occurs around the location where F1 rises up (e.g., the high temperature plasma shown in $\sim$ 01:42-01:52 UT in the bottom of Figure~\ref{stackplot}), it is reasonable to presume that part of the eruptive flux rope has high temperature and contributes to the formation of the high temperature plasma cloud that is beyond the temperature response in 304 \AA\  and is most obvious in high temperature response channel 131 \AA. 

During the same time period as the plasma moves from line cc to cd (corresponding to two crosses along curve C2 marked in Figure~\ref{filament}) in Figure~\ref{aiastack}, we observe high-temperature plasma drifting along curve C1 with speed $\sim$ 59 km/s, as indicated by S2 in the position-vs.-time stackplots of temperature in Figure~\ref{stackplot}. Given the relative locations of curves C1 and C2 and the geometry of the flare (see Figure~\ref{filament}) and the times of plasma movements marked by S1 (see Figure~\ref{aiastack}) and S2, we can reasonably presume that the drift of high-temperature plasma is related to ejected outflows from brightening intersection of eruptive filament F1 and pre-existing fialemnt F2 that is observed in 171 \AA\ and 304 \AA. This relationship is further confirmed by the fact that the plasma above the southern flare loops, which is quite far from filament brightening and ejected outflows, does not show noticeable plasma heating in Figure~\ref{T_aia_evolution}. 

We do not have direct evidence that the ejected outflows are the outflows of magnetic reconnection. However, combination of the facts that AIA 304 \AA\ and 171 \AA\ are two channels used to display inflows and outflows of magnetic reconnection in \citet{2014ApJ...788...85V}, comparable outflow speed (we have more projection effects than their case), additional heating of plasma cloud (S2 in Figure~\ref{stackplot}) geometrically located above the ejected outflows, and plasma above the southern flare loops shows no noticeable plasma heating, magnetic reconnection plausibly occurs between F1 and F2, leading to additional heating in the plasma cloud. In fact, the magnetohydrodynamic (MHD) simulations of \citet{2022ApJ...933..148C} and \citet{2024ApJ...977L..26C} suggest that magnetic reconnection occurs between the newly emerging flux and the erupting flux rope, resulting in the plasma heating around the reconnection site. The plasma related to temperature drift (indicated by S2 in Figure~\ref{stackplot}) in our case has the highest temperature among the plasma cloud (higher than 12 MK). 

\begin{figure*}[ht]
\centering
\includegraphics[scale=0.7]{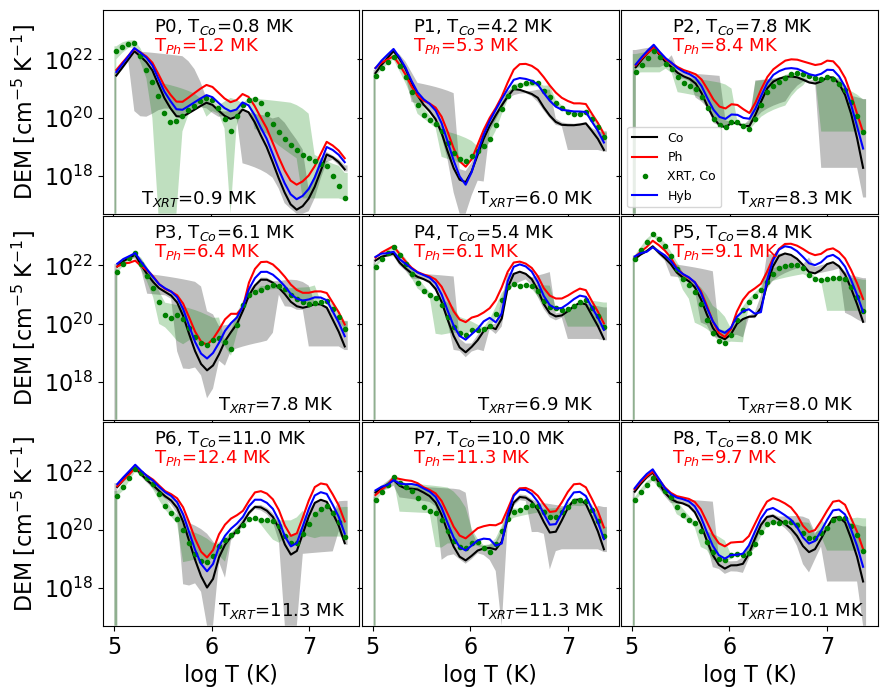}
\caption{DEM-temperature distributions at 01:55:19 UT in the locations marked as crosses in Figure~\ref{T_aia_evolution}. Solid curves indicate DEMs derived from AIA channels with coronal (black), hybrid (blue), and photospheric (red) abundance models. Green dotted curves indicate DEMs derived from the combination of XRT and AIA channels with the coronal abundance model. The uncertainties of black solid and green dotted curves are indicated by the same colored shadows. The uncertainties of red and blue curves are shown in Figure~\ref{peak_curve_appendix} of the Appendix to avoid confusion.} The DEM-weighted average temperature with coronal and photospheric abundance model for AIA-only case (T$_{Co}$ and T$_{Ph}$) and combining additional XRT filters (T$_{XRT}$) at each location is indicated in each panel.
\label{peak_diffabun_curve}
\end{figure*}

We further examine DEMs at 9 locations indicated by crosses shown in Figure~\ref{T_aia_evolution}. The DEMs with coronal (black), photospheric (blue), and hybrid (red) abundance models are shown in Figure~\ref{peak_diffabun_curve}. The curves among nine locations generally show three major DEM components. There is the commonly highest DEM value around temperature Log$_{10}$~T (K)~$\sim$5.2 coming from the emission of the background along the LOS. There is usually an apparent gap between the background and the temperature higher than 1~MK from the emission of the flare, except for the location P0 which corresponds to the non-flaring structure. For the locations in flaring structures (P1-P8), there are generally two DEM peaks around temperature Log$_{10}$~T (K)~$\sim$ 6.5 and Log$_{10}$~T (K)~$\sim$7.1. For the DEMs at the locations in the flare that do not show apparent heating (P1, P3, and P4), peaks in the DEMs around Log$_{10}$~T (K)~$\sim$6.5 are dominant for the temperature range higher than 1 MK. Compared to P1, P3, and P4, locations at heating loops (P2 and P5) and heated plasma cloud (P6-P8) illustrate much more contribution from high temperature range around Log$_{10}$~T (K)~$\sim$7.1. The DEM values for the locations at heated plasma cloud (P6-P8) around Log$_{10}$~T (K)~$\sim$7.1 are higher or comparable to Log$_{10}$~T (K)~$\sim$6.5, leading to high DEM-weighted average temperature $\overline{T}$ distribution in the plasma cloud (Figure~\ref{T_aia_evolution}). 

\subsection{Comparison of DEMs between different abundance models} 
\begin{figure}[h!]
\centering
\includegraphics[scale=0.5]{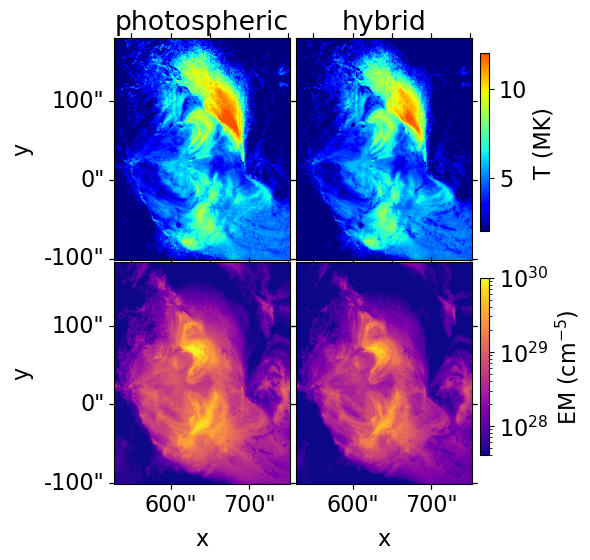}
\caption{Temperature and EM distributions at 01:55:19 UT derived from AIA data. We apply photospheric (left) and hybrid (right) abundance models for the comparison with the third column in Figure~\ref{T_aia_evolution}.}
\label{peak-diffabun}
\end{figure}
As a comparison, we calculate the temperature and EM distributions at 01:55 UT by applying photospheric and hybrid abundance models, as shown in Figure~\ref{peak-diffabun}. Compared to the coronal abundance assumption, the EM overall increases with the photospheric abundance assumption, and the EM with the hybrid abundance assumption has a value in between. The highest EM with photospheric abundances originates from that DEMs with photospheric abundance have the highest values, as displayed in Figure~\ref{peak_diffabun_curve}. Among the different abundance models, photospheric and coronal abundances correspond to the highest and lowest DEM values, respectively, which is expected from the temperature responses (e.g., Figure~\ref{tresponse}) and Equation~\ref{flux}. From Figure~\ref{peak-diffabun}, we note that the temperature distribution with photospheric abundance is generally slightly higher than that with coronal abundance. This higher DEM weighted average temperature with photospheric abundances is more specifically demonstrated in the DEMs at 9 locations (Figure~\ref{peak_diffabun_curve}).

\begin{figure*}[ht]
\centering
\includegraphics[scale=0.65]{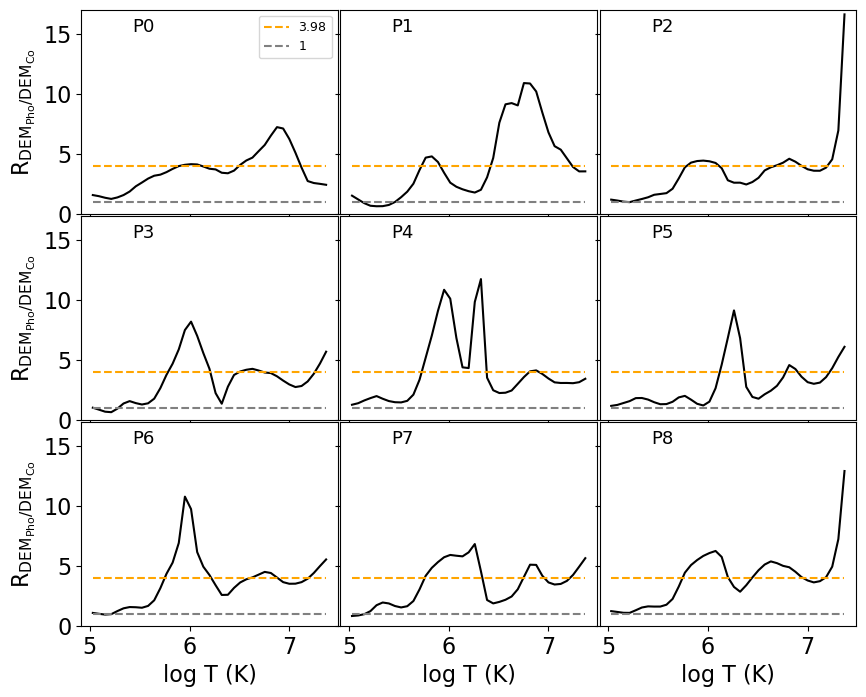}
\caption{Ratio of DEMs with photospheric abundances to DMEs with coronal abundances versus temperature at 01:55:19 UT in the locations marked by crosses in Figure~\ref{peak_diffabun_curve}. Grey and orange dashed lines indicate values equaling 1 and 3.98, respectively.}
\label{ratio}
\end{figure*}

We further assess the change of DEMs between photospheric and coronal abundances by plotting R$_{DEM_{pho}/DEM_{co}}$ (T), i.e., the ratio of DEMs with photospheric abundance to DEMs with coronal abundance versus temperature, for the nine selected locations in Figure~\ref{ratio}. R$_{DEM_{pho}/DEM_{co}}$ almost entirely $>$ 1 in the whole temperature range for all the checked locations. This ratio indicates that applying a single set of abundances over an entire image can underestimate the EM difference between features where coronal composition plasma via direct heating (e.g., plasma sheets in \citealp{2018ApJ...854..122W} or a plasma cloud as in the current paper) and photospheric/chromospheric composition plasma, such as flare loops formed via chromospheric evaporation, are both present. For some locations, i.e., p1, p2,p4, p6, and P8, R$_{DEM_{pho}/DEM_{co}}$ in certain temperature ranges is even higher than 10.

Also, Figure~\ref{ratio} demonstrates that compared with coronal abundances, implementing photospheric abundances results in a relatively higher DEM increase in temperatures $>$ 1~MK than in temperatures $<$ 1~MK. This relatively higher DEM increase leads to higher $\overline{T}$ with the photospheric abundance model, as shown in the comparison between Figures~\ref{T_aia_evolution} at 01:55:19 and Figure~\ref{peak-diffabun}. Remarkably, for the locations P2 and P5 that correspond to bright emission of flare loops, R$_{DEM_{pho}/DEM_{co}}$ is higher than 5 (even higher than 15 for P2) at temperatures $>$ 10 MK. This higher R$_{DEM_{pho}/DEM_{co}}$ indicates that in the locations where photopsheric abundance is supposed to apply, such as the locations with chromopsheric evaporation, implementing coronal abundances would underestimate (by a factor of 15 in some locations) DEMs in the high temperature range. 

R$_{DEM_{pho}/DEM_{co}}$ (T) also shows that the change in DEMs from a coronal abundance model to a photospheric abundance model is location-dependent. Therefore, compared to the coronal abundance model, implementing the photospheric abundance model does not simply scale the values of EM and temperature, and it leads to different EM and temperature distributions between different abundance models. The difference of EM or $\overline{T}$ with different abundance models suggests that instead of using the coronal abundance model by default to calculate DEMs throughout flares, it is necessary to apply appropriate abundance models to calculate DEMs in different phases/regions of flares from spectral diagnostics when it is applicable.

\subsection{DEMs combining XRT filters with AIA channels} 
\begin{figure*}
\centering
\includegraphics[scale=0.7]{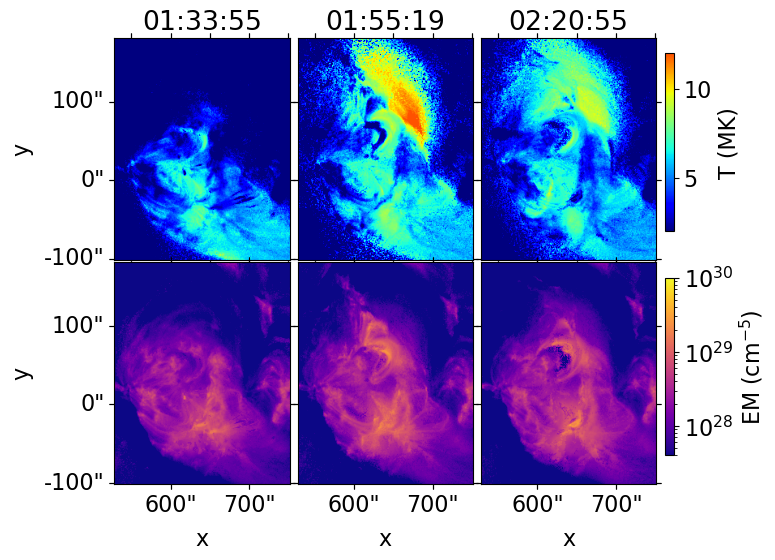}
\caption{Evolution of the temperature (top row) and EM (bottom row) from DEMs that combine the XRT and AIA channels.}
\label{T_mixture_evolution}
\end{figure*}

We further include XRT Be-med and Be-thin filters with AIA 6 channels to derive the DEMs. The results in Figure~\ref{T_mixture_evolution} show the consistent thermal evolution of the flare as in Figure~\ref{T_aia_evolution}, i.e., the flare loops heat up from impulsive phase to the flare peak and there is a heated higher temperature plasma cloud with lower density than the flare loops above the northern flare loops at 01:55. The high temperature plasma cloud and flare loops undergo different time scales of heating and cooling. The distributions of the temperature and EMs in Figure~\ref{T_mixture_evolution} are similar to the one in Figure~\ref{T_aia_evolution} while the temperature is overall slightly higher and the EMs are overall lower (note that, compared to Figure~\ref{T_aia_evolution}, we narrow the EM plotting colorbar range in Figure~\ref{T_mixture_evolution}) after we include two XRT filters to AIA 6 channels. 

We further examine DEMs at 01:55:19 UT for the case including XRT filters, shown as dotted curves in Figure~\ref{peak_diffabun_curve}. For 9 DEMs that represent typical locations, we see generally higher $\overline{T}$ after we include XRT filters. One exception is DEM at location P5 where we see apparent lower weights of DEMs above $\sim10^{6.5}$~K that results in lower $\overline{T}$ in the case of including XRT filters. The values can be either higher (i.e., P0-P2) or lower (P3-P8), but flatter and broader around $\sim10^{6.5}$~K after we include two XRT filters. The change of DEM shape around $\sim10^{6.5}$~K originates from a broad temperature response peak around $10^{6.5}$~K in XRT Be-med and Be-thin filters (Figure~\ref{tresponse}).

The relative DEMs above $\sim10^{7}$~K before/after adding XRT filters are location-dependent. We note, though, except for location P0, which corresponds to background non-flaring plasma, DEM uncertainties in the temperature $\gsim 10^{7}$~K generally decreases after we add XRT filters. 
Including instruments like MinXSS-1 and XRT, which cover a broader high-temperature range ($\gtrsim 10^{7.5}$~K) compared to AIA, reduces DEM uncertainties in the high-temperature ($\gtrsim 10^{7}$~K) regime. Moreover, \citet{2019ApJ...881..161M} demonstrate that the DEMs by combining the broadband spectrometer Reuven Ramaty High Energy Solar
Spectroscopic Imager (RHESSI; \citealp{2002SoPh..210....3L}) with EUV Variability Experiment (EVE; \citealp{2012SoPh..275..115W}) on the SDO better resolve ambiguities caused by energy overlap of thermal and nonthermal emissions, therefore better constraining the low-energy cutoffs of non-thermal electron distributions \citep{2011SSRv..159..107H}.
\section{Conclusion \& Discussion} 
\label{sec:discussion}
The statistical study on the abundance evolution of low FIP elements in \citet{2023ApJ...957...14S} shows the depletion of the abundances from coronal values before the eruptions toward photospheric values during the peaks in most flares, consistent with the chromospheric evaporation model. In contrast to the general scenario, the elemental abundances of the 2016-07-21 M1.0 flare (Figure~10 in \citealp{2023ApJ...957...14S}) shows an anomalous elemental abundance evolution, where the abundances of low-FIP elements increase from photospheric values in the impulsive phase to coronal values around the flare peak. This anomalous elemental abundance evolution motivates our initial interest in performing thermal diagnostics using the observations from XRT and AIA to investigate the source responsible for the anomalous elemental abundance evolution in the 2016-07-21 M1 flare. We note that there is a previous M2 flare that happened in the same active region as our studied flare. Chromospheric evaporation likely happened during this previous M2 flare, and it could explain the photospheric abundances observed at the beginning of our studied flare, as shown in the right panel of Figure~\ref{goes} and \citet{2023ApJ...957...14S}.

We find that the evolution of temperature and EM shows the accumulation of dense flare loops with high temperatures after the onset of the flare, presumably due to chromospheric evaporation. After the flare impulsive phase, the filament flux rope F1 rises up, and a tenuous plasma cloud, which is most obviously observed in the AIA 131~\AA\ channel, is located above the rising filament that is observed in the AIA 304~\AA\ channel. The formation of the plasma cloud in 131 \AA\ is likely a heated eruptive flux rope that is not visible in the low-temperature channel of 304 \AA\ and is detected by the hot temperature channel of 131~\AA.

When the eruptive flux rope F1 encounters the pre-existing filament F2, we observe ejected outflows with the plane-of-sky speed of 44 km/s from the intersection of F1 and F2 in AIA 304 \AA\ and 171 \AA. At the same time, there is additional plasma heating in the plasma cloud geometrically located above the ejected outflows. We do not have direct evidence that the ejected outflows are the outflows of magnetic reconnetion. However, combination of the facts that the speeds we observe in AIA 304 \AA\ and 171 \AA\ are comparable to the inflows and outflows of magnetic reconnection observed by \citet{2014ApJ...788...85V}, and additional heating of plasma cloud geometrically located above the ejected outflows, magnetic reconnection plausibly occurs between F1 and F2. This conclusion is further confirmed by that the plasma above the southern flare loops, which are far from ejected outflows, shows no noticeable plasma heating. The temperature distribution shows that the area of the plasma cloud with additional heating has the highest temperature around the flare peak. The heated plasma cloud is dominant for highest temperature around the flare peak.

The heating of the plasma cloud around the flare peak derived from AIA and XRT data could explain coronal abundances of the low FIP elements in \citet{2023ApJ...957...14S}. The flare loops and the plasma cloud show independent heating and cooling process, the highest temperature sequentially evolves from flare loops to plasma cloud and the long duration of high temperature plasma cloud contributes to long duration ($\sim$1 hour) of high temperature ($\gsim$ 7~MK) throughout the impulse, peak, and decay phases of the flare diagnosed in \citet{2023ApJ...957...14S}. A plausible picture where plasma cloud has coronal composition and undergoes direct plasma heating in the corona around the flare peak is thus given. 

However, MinXSS-1 synthetic spectra from AIA observations shows that the plasma cloud is too tenuous to be detected in MinXSS-1. Therefore, we can not exclude that the coronal abundances of low FIP elements come from flare loops \citep{2024A&A...691A..95T}. It requires more future studies to investigate the mechanisms responsible for coronal abundance detected at the flare peak.

The details of the evolution of the abundance variation in eruptions at relatively small scales is likely quite complex. For example, \citet{2018ApJ...853..178D} and \citet{2024A&A...691A..95T} report FIP bias variation along the flare loops, possibly due to a combination of confinement of high FIP bias reconnection downflows from the plasma sheet in the loop tops and chromospheric evaporation that fills the loop footpoints from below with low FIP bias plasma, or a blend of loops or magnetic strands along the line of sight. Even for the statistical studies that illustrate the chromospheric evaporation from whole Sun spectra, other heating processes resulting in other compositions could occur during the flare, and the information could be hidden in whole Sun spectral diagnostics.

In addition, we illustrate that compared to DEM calculations with only AIA channels, including additional XRT Be-med and Be-thin broadband filters leads to generally higher $\overline{T}$ and broader DEM curves around $10^{ 6.5}$ K due to a broad temperature response around $10^{6.5}$~K in XRT Be-med and Be-thin filters. The change of DEMs above $10^7$~K before/after including XRT filters is location-dependent. We found that adding Be-med and Be-thin filters better constrains DEMs in  high-temperature ($\gtrsim 10^{7}$~K) regime. 

In the future, instruments with the capability of simultaneous spectral analysis and spatial resolution, such as the potential  Scotopic Solar Activity X-ray Imager; SSAXI-SmallSat \citep{Hong2019}, the Swift Solar Activity X-ray Imager Rocket; SSAXI-Rocket \citep{Moore2024}, a satellite version of Focusing Optics X-ray Solar Imager; FOXSI \citep{Krucker2014, Christe16, Glesener2016}, and the Marshall Grazing Incidence X-ray Spectrometer 2 (MaGIXS-2) \citep{2025arXiv250814866M} will be useful to identify plasma source locations, diagnose plasma composition, and investigate physical processes during large-scale solar eruptions simultaneously. Such instruments are crucial because they provide elemental abundances at different flare locations and phases. This need is further demonstrated in our results showing that the difference between DEMs with different abundance models is location-dependent and non-negligible (DEMs increase as much as 15 times for certain temperature ranges at some locations from coronal abundances to photospheric abundances). 
 
\section*{Acknowledgments}
We are grateful to Dr. William Ashfield for the suggestion and help for generating AIA response functions. We thank the anonymous referee for comments and suggestions that improved the paper. The work of X. X. and C. S. is supported by contract NNM07AB07C from NASA to SAO. K.K.R.and S.R. are supported by NASA HSO Connect grant 80NSSC20K1283. C.S. is also supported by the NASA-HSR NSSC20K1447 and Vanderbilt Bridge Program PhD Fellowship. The work of  A.C. was supported by grant number AGS-2244112 and the NSF-REU solar physics program at SAO. The MinXSS-1 CubeSat mission was supported by NASA grant NNX14AN84G. The AIA data are provided courtesy of NASA/SDO and the AIA science team. Hinode is a Japanese mission developed and launched by ISAS/JAXA, with NAOJ as domestic partner and NASA and STFC (UK) as international partners. It is operated by these agencies in co-operation with ESA and NSC (Norway).  The SolarSoftWare (SSW) system is built from Yohkoh, SOHO, SDAC and Astronomy libraries and draws upon contributions from many members of those projects. CHIANTI is a
collaborative project involving George Mason University,
the University of Michigan (USA), and the University of
Cambridge (UK). We would also like to extend our thanks to the developers of XRTpy: A Python package for solar X-ray telescope data analysis (available at https://xrtpy.readthedocs.io/) for providing the tools necessary for our XRT analysis.
\appendix \label{sec:app}
\renewcommand\thefigure{\thesection.\arabic{figure}}
\setcounter{figure}{0} 
\section{DEM uncertainties}
\begin{figure*}[h!]
\centering
\includegraphics[scale=0.6]{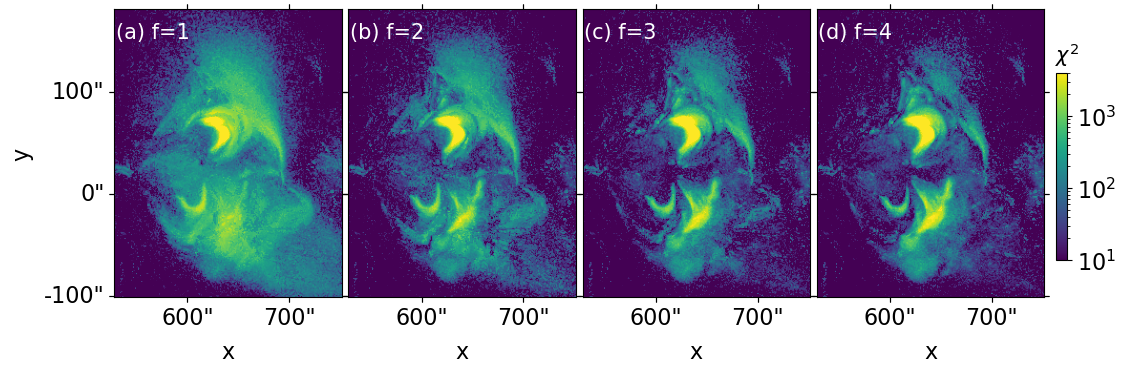}
\caption{$\chi^2$ distributions with different factors multiplying the XRT response function when combining XRT filters with AIA channels to calculate DEMs at 01:55:19 UT.}
\label{xrtfactors}
\end{figure*}

\begin{figure*}[ht!]
\centering
\includegraphics[scale=0.7]{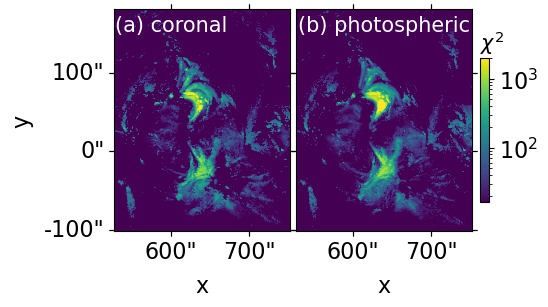}
\caption{$\chi^2$ with coronal and photospheric models at 01:55:19 UT. Panel (a) corresponds to Figure~\ref{T_aia_evolution} at 01:55:19 UT, and panel (b) correspond to left column of Figure~\ref{peak-diffabun}. It shows that, for loops, $\chi^2$ with coronal abundances is generally lower than that with photospheric abundances.}
\label{chisq}
\end{figure*}
\begin{figure*}[ht!]
\centering
\includegraphics[scale=0.7]{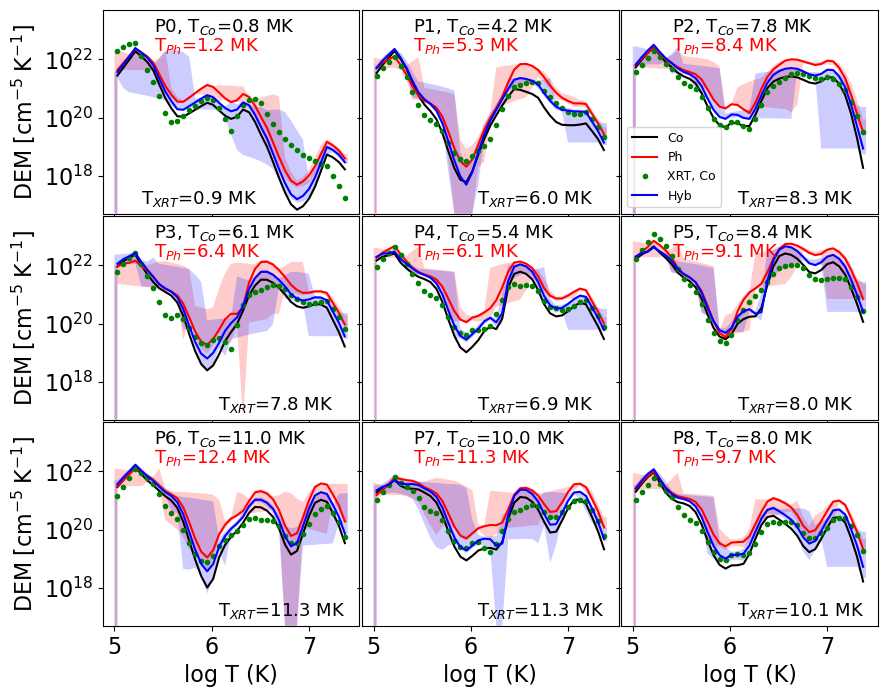}
\caption{Same as Figure~\ref{peak_diffabun_curve} but displaying the DEM uncertainties with photospheric and hybrid abundance models.}
\label{peak_curve_appendix}
\end{figure*}
Here, we show DEM uncertainties to support our conclusions in the paper. Figure~\ref{xrtfactors} shows $\chi^2$ distributions with different factors multiplying  the XRT response function when combining XRT filters with AIA channels to calculate DEMs. Comparing using a multiplying factor of 1 for XRT response function (i.e., f=1 in Figure~\ref{xrtfactors}a), multiplying XRT response function by a factor of 2 generally decreases  $\chi^2$.  Figure~\ref{xrtfactors} also illustrates that increasing the multiplying factor from 2 to 3 or 4 does not generally decrease $\chi^2$, and $\chi^2$ distributions in the loops increase with multiplying factors of 3 or 4. We therefore multiply XRT response function by a factor of 2 in our DEM calculations of combining XRT filters with AIA channels.

Figure~\ref{chisq} shows that loops with coronal abundances have generally lower $\chi^2$ than that with photospheric abundances, supporting the existence of coronal abundances in flare loops. Figure~\ref{peak_curve_appendix} is the complementary figure to Figure~\ref{peak_diffabun_curve}. We show the DEM uncertainties with photospheric and hybrid abundance
models in Figure~\ref{peak_curve_appendix} to avoid confusion in Figure~\ref{peak_diffabun_curve}.
\section{MinXSS synthetic spectra from AIA observations}
\setcounter{figure}{0} 
\begin{figure}[ht!]
\centering
\includegraphics[scale=0.5]{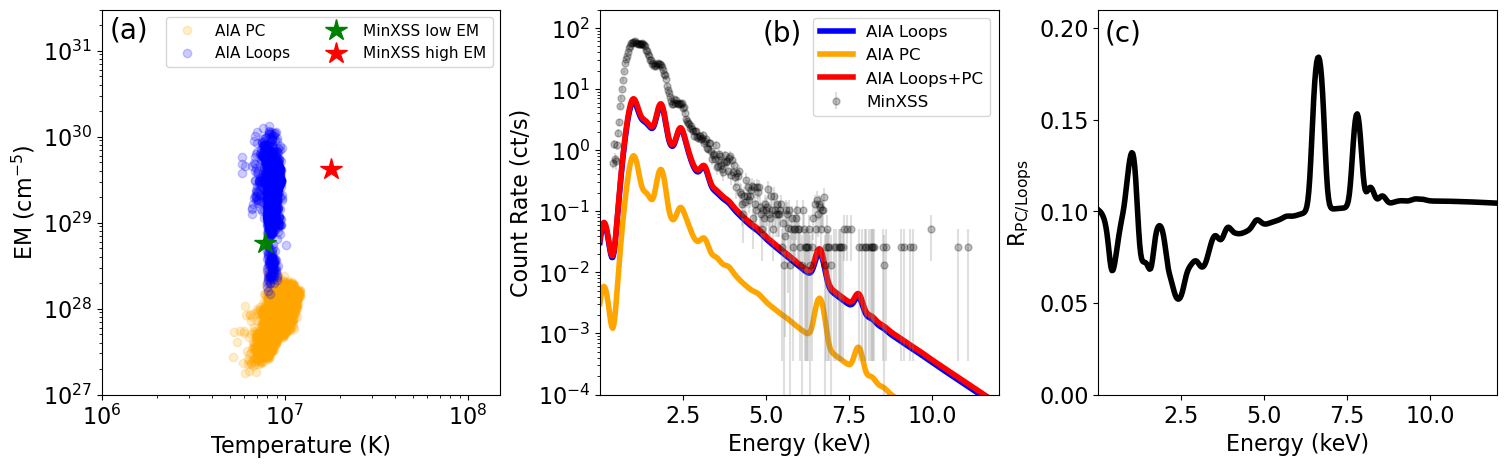}
\caption{(a): EM versus temperature distribution for plasma cloud and loops in AIA where MinXSS low and high EM versus temperature are indicated by stars. (b): Synthetic spectra MinXSS$_{\rm AIA}$ of loops, plasma cloud, and flare (loop+plasma cloud) from AIA observations. As a reference, MinXSS observations are shown in (b). (c): Ratio of plasma cloud synthetic spectrum to the loop synthetic spectrum.}
\label{syn}
\end{figure}

Since AIA conducts spatially resolved measurements and MinXSS is disk-integrated spectrometer without spatial resolution, we calculate the MinXSS synthetic spectra from AIA DEM calculations to assess whether the tenuous plasma cloud could be detected in MinXSS. We first apply masks to select the pixels of plasma cloud and loops separately and plot the EM versus temperature for each pixel of plasma cloud and loops, as shown in Figures~\ref{syn}(a). 
The EM distribution of the loops is overall higher than that of plasma cloud. We also plot the corresponding MinXSS low/high EM versus temperature with the assumption that MinXSS spectra dominantly come from the flare. Interestingly, the low MinXSS derived EM and T are located in the AIA loops and the high MinXSS derived EM and temperature is outside of the AIA EM versus temperature distribution of the whole flare. 
 
We pass the temperature and EM derived from AIA through the MinXSS instrument response (CHIANTI 10, \citealp{1997A&AS..125..149D,2021ApJ...909...38D}) to obtain synthetic MinXSS$_{\rm AIA}$ for the plasma cloud and loops (Figure~\ref{syn}b). A similar practice of the MinXSS synthetic spectra can be found in \citet{2020ApJ...895...30R}. The MinXSS observations are plotted in Figures~\ref{syn}(b) as well. We note that the majority of the synthetic MinXSS spectral signal is from the AIA loops (blue curve in Figure~\ref{syn}b). Figure~\ref{syn}(b) and the ratio of plasma cloud synthetic
spectrum to the loop synthetic spectrum in Figure~\ref{syn}(c) shows that the synthetic spectrum by plasma cloud is about one order of magnitude lower (Figure~\ref{syn}b, orange curve) than that by loops (blue curve, which makes the difference between synthetic spectra of plasma cloud and flare (plasma cloud+loops) in Figure~\ref{syn}(b) almost unnoticeable. 

\section{Impact of filling factor}
Calculation of the number density $n$ from EM is via $n=\sqrt{{\rm EM}/f\rm L}$, where $f$ is a filling factor that characterizes the occupying fraction of emitting plasma to the apparent volume and $L$ is the depth along LOS. Here, we assess whether the value of $f$ would alter our findings in a meaningful way. We adopt the value of $f$ specified in Section 3.3.4 of \citet{2020A&A...644A.172W}, i.e., $ 0.1 < f < 1$, as the possible filling factors of the plasma cloud and loops. This range is consistent with the median value of $f$ = 0.12 reported by \citet{2021ApJ...906...59H}, based on measurements from 20 coronal bright points with available spectral data. The ratio of filling factor from plasma cloud to loops then is $ 0.1 < R_{f_{\rm PC}/f_{\rm loops}} < 10$. From EMs that correspond to Figures~\ref{syn}(a), we obtain the ratio of plasma cloud EM median to loop EM median as $ R_{\rm EM_{\rm PC}/\rm EM_{\rm loops}}$=0.029. Assuming L is the same for the plasma cloud and loops, the ratio of plasma cloud number density to loop number density is $n_{\rm PC}/n_{\rm loops} = \sqrt{ \frac{R_{\rm EM_{\rm PC}/\rm EM_{\rm loops}}} {R_{f_{\rm PC}/f_{\rm loops}}}}$, which is in the range from 0.054 to 0.54. The range of $n_{\rm PC}/n_{\rm loops}$ suggests that the main conclusion in the paper, i.e., the loops are denser and the plasma cloud is tenuous, which could lead to the plasma cloud undetectable in MinXSS-1, still holds from number density perspective with the impact of the filling factor included.


\bibliography{sample631}{}
\bibliographystyle{aasjournal}



\end{document}